\documentclass[aps,pra,twocolumn,floatfix]{revtex4}

\usepackage{latexsym}
\usepackage{amsmath}
\usepackage{graphicx}
\usepackage{float}

\begin{document}

\title{Magnetic and thermal properties of a one-dimensional spin-1 model}
\author{F. Mancini}
\affiliation{ Dipartimento di Fisica ``{\it E. R. Caianiello}'' -
Unit\`a CNISM di Salerno \\ Universit\`a degli Studi di Salerno,
Via S. Allende, I-84081 Baronissi (SA), Italy}

\author{F. P. Mancini}
\affiliation{ Dipartimento di Fisica and Sezione I.N.F.N.
\\
Universit\`a degli Studi di Perugia, Via A. Pascoli, I-06123
Perugia, Italy}

\begin{abstract}
We study the one-dimensional $S=1$ Blume-Emery-Griffiths model.
Upon transforming the spin model into an equivalent fermionic
model, we provide the exact solution within the Green's function
and equations of motion formalism. We show that the relevant
response functions as well as thermodynamic quantities can be
determined, in the whole parameters space, in terms of a finite
set of local correlators. Furthermore, considering the case of an
antiferromagnetic chain with single-ion anisotropy in the presence
of an external magnetic field, we find three plateaus in the
magnetization curve; in the neighborhood of the endpoints of the
intermediate plateau, the specific heat shows a double peak
structure.

\end{abstract}

\date{\today}
\maketitle

\section{Introduction}

The spin-1 Ising model with bilinear ($J$) and biquadratic ($K$)
nearest-neighbor pair interactions and a single-ion potential
($\Delta$) is known as the Blume-Emery-Griffiths (BEG) model
\cite{blume71}. With vanishing biquadratic interactions, the model
is known as the Blume-Capel model \cite{blume66,capel6667}.

There is a large and diffused interest in the study of this model,
motivated by several reasons. The BEG model was originally
introduced to describe the phase separation and superfluidity in
the $^3$He-$^4$He mixtures, but it can also describe the
properties of a variety of systems ranging from spin-1 magnets  to
liquid crystal mixtures, microemulsions, semiconductor alloys, to
quote a few. Both the BEG and BC models have been investigated
using many different approaches for different lattice type and
dimensions. In one dimension and zero magnetic field, the spin-1
Ising model and the BEG model have been solved exactly by means of
the transfer matrix method \cite{Suzuki67_Hintermann69} and by
means of the Bethe method \cite{Obokata68}. Exact solutions have
also been obtained for a Bethe lattice \cite{Chakraborty86} and
for the two-dimensional honeycomb lattice \cite{Rosengren89}. For
higher dimensions, among the various approximate and simulation
techniques, the most common approach to the BC and BEG models is
based on the use of mean field approximation. However,
renormalization group studies show some qualitative differences
from the mean field results. The one-dimensional case for the BEG
model was studied in Ref.~\cite{krinsky75}, where exact
renormalization-group recursion relations were derived, exhibiting
tricritical and critical fixed points. We refer the interested
reader to Ref. \cite{mancini05b} for a rather broad list of works
devoted to the study of the BC and BEG models.

In previous works \cite{mancini05s}, we have shown that, upon
transforming to fermionic variables, spin systems can be
conveniently studied by means of quantum field methods, namely:
Green's functions and equations of motion methods. This approach
has the advantage of offering a general formulation for any
dimension and to provide a rigorous determination of a complete
set of eigenoperators of the Hamiltonian and, correspondingly, of
the set of elementary excitations. In this paper we apply this
formulation to the 1$D$ BEG model, and we obtain the exact
solution of the model in the whole space of the parameters.
Postponing to a forthcoming work a comprehensive analysis of the
model, in this article we focus our study on the thermal and
magnetic properties of the antiferromagnetic ($J<0$) spin-1 chain
with single-ion anisotropy in the presence of an external magnetic
field, in the limit of vanishing biquadratic interaction. This
particular case is indeed interesting since one finds
magnetization plateaus, experimentally observed
\cite{experiments}. At $T=0$ the magnetization curve forms
plateaus with abrupt jumps from one to another at certain values
of the magnetic field $h$. When $\Delta=0$, the ground state is
purely antiferromagnetic for $-2\vert J \vert < h<2 \vert J
\vert$. By varying the magnetic field, the system undergoes a
phase transition to a pure ferromagnetic regime at $h= \pm 2 \vert
J \vert$. When the anisotropy $\Delta$ is turned on, one observes
an intermediate phase characterized by half of the spins oriented
along the external field and the rest perpendicular to it. The
intermediate phase between the anti- and ferromagnetic ones, has a
width depending on $\Delta$ whose endpoints are denoted by $h_c$
and $h_s$, i.e., critical and saturated field, respectively.

The paper is organized as follows. In Section \ref{sec_II}, upon
introducing a complete set of composite operators, eigenoperators
of the Hamiltonian, we outline the analysis leading to the algebra
closure and to analytical expressions of the retarded Green's
functions (GFs) and correlation functions (CFs). Since the
composite operators do not satisfy a canonical algebra, the GF and
CF depend on a set of internal parameters, leading only to exact
relations among the CFs. According to the scheme of the composite
operator method \cite{manciniavella}, it is possible to determine
these parameters by means of algebra constraints fixing the
representation of the GF. By following this scheme, in Section
\ref{sec_III} we obtain extra equations closing the set of
relations and allowing for an exact and complete solution of the
1$D$ BEG model.  Section \ref{sec_IV} is devoted to the study of
the finite temperature properties. Finally, Sec. \ref{sec_V} is
devoted to our conclusions and final remarks, while the appendix
reports some relevant computational details.

\section{Composite fields and Green's functions}
\label{sec_II}

The Blume-Emery-Griffiths (BEG) model consists of a system with
three states per spin. For first-nearest neighbor interaction the
one-dimensional BEG model is described by the Hamiltonian
\begin{equation}
\label{eq1}
\begin{split}
H &= -J\sum_i S(i)S(i+1)-K\sum_i S^2(i)S^2(i+1) \\
 &+\Delta \sum_i
S^2(i)-h\sum_i S(i),
\end{split}
\end{equation}
where the spin variable $S(i)$ takes the values $S(i)=-1,0,1$. We
use the Heisenberg picture:  $i=(i,t)$, where $i$  stands for the
lattice vector  $R_i$. This model can be mapped into a fermionic
model by means of the transformation
\begin{equation}
\label{eq2} S(i)=[n(i)-1],
\end{equation}
where $n(i)=\sum_\sigma c_\sigma ^\dag (i)c_\sigma (i)=c^\dag
(i)c(i)$ is the density number operator of a fermionic system;
$c(i)$ ($c^\dag (i))$ is the annihilation (creation) operator of
fermionic field in the spinor notation and satisfies canonical
anti-commutation relations. Under the transformation (\ref{eq2}),
 the Hamiltonian (\ref{eq1})
takes the form
\begin{equation}
\label{eq3}
\begin{split}
 H &=V\sum_i n(i)n^\alpha (i)+\tilde {U}\sum_i D(i)
\\
 &+\frac{1}{2}W\sum_i
n(i)D^\alpha (i) +\frac{1}{2}W\sum_i D(i)n^\alpha (i) \\
&-W\sum_i D(i)D^\alpha (i)-\tilde
 {\mu}
\sum_i n(i)+E_0 ,
 \end{split}
\end{equation}
where we have defined
\begin{equation}
\label{eq4}
\begin{split}
V&=-(J+K), \\
\tilde {U}&=2(-2K+\Delta ), \\
W&=4K,
\end{split}
\quad
\begin{split}
\tilde {\mu } &=-(2J+2K-\Delta -h), \\
E_0 &=(-J-K+\Delta +h)N .
\end{split}
\end{equation}
Hereafter, for a generic operator $\Phi(i)$ we shall use the
following notation: $\Phi^{\alpha}(i)=[ \Phi(i+1)+\Phi(i-1) ]/2$.
The Hamiltonian \eqref{eq3} is pertinent to an Hubbard model
extended to include intersite interactions, namely: charge-charge
($V$), charge-double occupancy ($W$) and double occupancy-double
occupancy ($-W$) interactions. However, Hamiltonian (\ref{eq3})
does not exactly corresponds to the BEG Hamiltonian (\ref{eq1})
since the mapping between $S$ and $n$ should take into account the
four possible values of the particle density ($n(i)=0$,
$n_{\uparrow}(i)=1$ and $n_{\downarrow}(i)=1$, $n(i)=2$). Letting
the zero-state spin being degenerate, makes the Hamiltonians
\eqref{eq1} and \eqref{eq3} equivalent, provided one redefines the
chemical potential $\tilde{\mu}$ and the on-site potential
$\tilde{U}$ as
\begin{equation}
\label{eq5}
\begin{split}
 \tilde {\mu } &\to \mu =\tilde{\mu} -\beta ^{-1}\ln 2, \\
\tilde {U}  &\to U=\tilde {U}-2\beta ^{-1}\ln 2 ,
 \end{split}
\end{equation}
where $\beta =1/k_B T$. As a result, for a translationally
invariant chain, one has
\begin{equation}
\label{eq6}
\begin{split}
H&=V\sum_i n(i)n^\alpha (i)+U \sum_i D(i)+W\sum_i n(i)D^\alpha (i)
\\
&-W\sum_i D(i)D^\alpha (i)-\mu \sum_i n(i)+E_0.
 \end{split}
 \end{equation}
It is easy to see that the partition functions relative to the two
models (\ref{eq1}) and (\ref{eq6}) are the same. Therefore, in the
following we shall consider the fermionic model, described by the
Hamiltonian (\ref{eq6}), which is exactly equivalent to the BEG
model. To solve this Hamiltonian we shall use the formalism of
Green's functions and equations of motion. As a first step, we
show that there exists a complete set of eigenoperators and
eigenvalues of $H$. To this purpose, one can introduce the Hubbard
projection operators
\begin{equation}
\label{eq7}
\begin{split}
 \xi (i)&=[1-n(i)]c(i), \\
 \eta (i)&=n(i)c(i).
 \end{split}
\end{equation}
These fields satisfy the equations of motion
\begin{equation}
\label{eq8}
\begin{split}
 i\frac{\partial }{\partial t}\xi (i)&=\left[-\mu +2Vn^\alpha
(i)+WD^\alpha (i)\right]\, \xi (i), \\
 i\frac{\partial }{\partial t}\eta (i)&=\left[-\mu +U+(2V+2W)n^\alpha
(i)-WD^\alpha (i)\right]\, \eta (i).
 \end{split}
\end{equation}
By noting that the particle density $n(i)$ and double occupancy
$D(i)$ operators satisfy the following algebra
\begin{equation}
\label{eq9}
\begin{split}
n^p(i)&=n(i)+a_p D(i), \\
D^p(i)&=D(i), \\
n^p(i)D(i)&=2D(i)+a_p D(i),
\end{split}
\quad \quad {\rm for} \quad p\ge 1,
\end{equation}
where $a_p =2^p-2$, it is easy to derive the following recursion
rules
\begin{equation}
\label{eq10}
\begin{split}
[n^\alpha (i)]^p&=\sum_{m=1}^4 A_m^{(p)} [n^\alpha (i)]^m ,\\
 [D^\alpha (i)]^p&=\sum_{m=1}^2 B_m^{(p)} [D^\alpha
(i)]^m .
\end{split}
\end{equation}
The coefficients $A_m^{(p)} $ and $B_m^{(p)}$ are rational
numbers, satisfying the sum rules $\sum_{m=1}^4 A_m^{(p)} =1$ and
$\sum_{m=1}^2 B_m^{(p)} =1$, whose explicit expressions are given
in the appendix.

On the basis of the equations of motion (\ref{eq8}) and by means
of the recursion rules \eqref{eq10}, it is easy to see that the
composite multiplet operators
\begin{equation}
\label{eq14} \psi ^{(\xi )}(i)=\left( {{\begin{array}{*{20}c}
 {  \scriptstyle \xi (i)}  \\
 {  \scriptstyle \xi (i)n^\alpha (i)}  \\
 {  \scriptstyle \xi (i)[n^\alpha (i)]^2}  \\
 {  \scriptstyle \xi (i)[n^\alpha (i)]^3}  \\
 {  \scriptstyle \xi (i)[n^\alpha (i)]^4}  \\
 {  \scriptstyle \xi (i)D^\alpha (i)}  \\
 {  \scriptstyle \xi (i)[D^\alpha (i)]^2}  \\
\end{array} }} \right), \; \psi ^{(\eta )}(i)=\left(
{{\begin{array}{*{20}c}
 {  \scriptstyle \eta (i)}  \\
 {  \scriptstyle \eta (i)n^\alpha (i)}  \\
 {  \scriptstyle \eta (i)[n^\alpha (i)]^2}  \\
 {  \scriptstyle \eta (i)[n^\alpha (i)]^3}  \\
 {  \scriptstyle \eta (i)[n^\alpha (i)]^4}  \\
 {  \scriptstyle \eta (i)D^\alpha (i)}  \\
 {  \scriptstyle \eta (i)[D^\alpha (i)]^2}  \\
\end{array} }} \right)
\end{equation}
are eigenoperators of $H$:
\begin{equation}
\label{eq15}
\begin{split}
i\frac{\partial }{\partial t}\psi ^{(\xi )}(i)&=[\psi ^{(\xi
)}(i),H]=\varepsilon ^{(\xi )}\psi ^{(\xi )}(i),
\\
 i\frac{\partial }{\partial t}\psi ^{(\eta )}(i)&=[\psi ^{(\eta
)}(i),H]=\varepsilon ^{(\eta )}\psi ^{(\eta )}(i).
\end{split}
\end{equation}
The energy matrices $\varepsilon^{(\xi )}$ and $\varepsilon
^{(\eta )}$ are matrices of rank $7\times 7$ and have the
expressions
\begin{widetext}
\begin{equation}
 \label{eq17}
 \varepsilon ^{(\xi)}=
 \left(
{  {\begin{array}{*{7}c}
 \scriptstyle {-\mu }  &  \scriptstyle {2V}  &  \scriptstyle 0  &  \scriptstyle 0  &  \scriptstyle
 \scriptstyle 0  &  \scriptstyle {W}
&  \scriptstyle 0
\\
\scriptstyle  0  &  \scriptstyle {-\mu +\frac{W}{6}}  &
\scriptstyle {2V-\frac{1}{2}W}  &  \scriptstyle {\frac{1}{3}W}
 &  \scriptstyle 0  &  \scriptstyle {W}  &  \scriptstyle 0
 \\
\scriptstyle  0  &  \scriptstyle {\frac{W}{6}}  &  \scriptstyle
{-\mu -\frac{1}{3}W}  &  \scriptstyle {2V-\frac{W}{6}}  &
\scriptstyle {\frac{1}{3}W}  &  \scriptstyle {W}  &  \scriptstyle
0
\\
\scriptstyle  0  &  \scriptstyle {-\frac{1}{3}W}  &  \scriptstyle
{\frac{7}{4}W}  &  \scriptstyle {-\mu -\frac{35}{12}W}  &
\scriptstyle {2V+\frac{3W}{2}}  &  \scriptstyle {W}  &
\scriptstyle 0
 \\
\scriptstyle  0  &  \scriptstyle {-3V-\frac{25}{12}W}  &
\scriptstyle {\frac{25}{2}V+\frac{205}{24}W}
 &  \scriptstyle {-\frac{35}{2}V-\frac{265}{24}W}  &  \scriptstyle {-\mu
+10V+\frac{55}{12}W}
 &  \scriptstyle {W}  &  \scriptstyle 0
  \\
\scriptstyle  0  &  \scriptstyle {\frac{1}{3}V}  &  \scriptstyle
{-V}  &  \scriptstyle {\frac{2}{3}V}  &  \scriptstyle 0
 &  \scriptstyle {-\mu +2V}  &  \scriptstyle {W}
 \\
\scriptstyle  0  &  \scriptstyle {-\frac{1}{3}V}  &  \scriptstyle
{\frac{4}{3}V}  &  \scriptstyle {-\frac{5}{3}V}
 &  \scriptstyle {\frac{2}{3}V}  &  \scriptstyle {-\frac{W}{2}}  &  \scriptstyle {-\mu +2V+\frac{3W}{2}}
 \\
\end{array} }} \right),
\end{equation}
\begin{equation}
 \varepsilon ^{(\eta )}=
\begin{pmatrix}
 \scriptstyle -(\mu -U)  &  \scriptstyle \Upsilon  & \scriptstyle 0  & \scriptstyle 0  &  \scriptstyle 0  &
 \scriptstyle -W  &  \scriptstyle 0
  \\
 \scriptstyle 0  &  \scriptstyle U-\mu -\frac{1}{6}W  &  \scriptstyle 2V+\frac{3W}{2}  &  \scriptstyle -\frac{1}{3}W
 &  \scriptstyle 0  &  \scriptstyle -W  &  \scriptstyle 0
  \\
 \scriptstyle 0  &  \scriptstyle -\frac{1}{6}W  &  \scriptstyle U-\mu+\frac{1}{3}W  &  \scriptstyle
2V+\frac{7}{6}W  &  \scriptstyle -\frac{1}{3}W  &  \scriptstyle -W
&  \scriptstyle 0
\\
 \scriptstyle 0  &  \scriptstyle \frac{1}{3}W  &  \scriptstyle -\frac{7}{4}W  &  \scriptstyle U-\mu
+\frac{35}{12}W &  \scriptstyle 2V-\frac{W}{2}  &  \scriptstyle -W
&  \scriptstyle 0
\\
 \scriptstyle 0  &  \scriptstyle \frac{7}{12}W-3V  &  \scriptstyle \frac{25}{2}V-\frac{55}{24}W
 &  \scriptstyle \frac{55}{24}W-\frac{35}{2}V  &  \scriptstyle U-\mu+10V+\frac{5}{12}W
 &  \scriptstyle -W &  \scriptstyle 0
  \\
 \scriptstyle 0  &  \scriptstyle \frac{1}{6}\Upsilon  &  \scriptstyle - \frac{1}{2}\Upsilon  &  \scriptstyle \frac{1}{3}\Upsilon
 &  \scriptstyle 0  &  \scriptstyle U-\mu +\Upsilon  &  \scriptstyle -W
 \\
   \scriptstyle 0  &  \scriptstyle -\frac{1}{6}\Upsilon  &  \scriptstyle \frac{2}{3}\Upsilon  &  \scriptstyle
-\frac{5}{6}\Upsilon  &  \scriptstyle \frac{1}{3}\Upsilon  &
\scriptstyle \frac{W}{2}  & \scriptstyle U-\mu +2V-\frac{1}{2}W
\end{pmatrix},
\end{equation}
\end{widetext}
 where $\Upsilon=2V+W$. The energy levels are given
by the eigenvalues of the energy matrices and are
\begin{equation}
\label{eq18} E_n^{(\xi )} =
\begin{pmatrix}
  \scriptstyle -\mu \\
  \scriptstyle -\mu +V \\
  \scriptstyle -\mu +2V \\
  \scriptstyle -\mu +W+2V \\
  \scriptstyle -\mu +W/2+2V \\
  \scriptstyle -\mu +W+4V \\
  \scriptstyle -\mu +W/2+3V \\
\end{pmatrix},
\; E_n^{(\eta)} =
\begin{pmatrix}
  \scriptstyle -\mu +U \\
  \scriptstyle -\mu +U+W/2+V \\
  \scriptstyle -\mu +U+W+2V \\
  \scriptstyle -\mu +U+2V \\
  \scriptstyle -\mu +U+W/2+2V \\
  \scriptstyle -\mu +U+W+4V \\
  \scriptstyle -\mu +U+W+3V \\
\end{pmatrix} .
\end{equation}
The knowledge of a complete set of eigenoperators and eigenvalues
of the Hamiltonian allows for an exact expression of the retarded
Green's function
\begin{equation}
\label{eq19}
\begin{split}
 G^{(s)}(t-t')&=\theta (t-t')\langle
\{\psi ^{(s)}(i,t),{\psi ^{(s)}}^\dag (i,t')\} \rangle
 \\
&= \frac{i}{(2\pi )}\int_{-\infty }^{+\infty } d\omega {\kern
1pt}e^{-i\omega (t-t')}G^{(s)}(\omega ),
\end{split}
\end{equation}
and, consequently, of the correlation function
\begin{equation}
\label{eq20}
\begin{split}
C^{(s)}(t-t')&=\langle \psi ^{(s)}(i,t){\psi ^{(s)}}^\dag
(i,t')\rangle \\
&=\frac{1}{(2\pi )}\int_{-\infty }^{+\infty } d\omega {\kern
1pt}e^{-i\omega (t-t')}C^{(s)}(\omega ).
\end{split}
\end{equation}
In the above equations $s=\xi ,\eta $ and $\langle \cdots \rangle$
denotes the quantum-statistical average over the grand canonical
ensemble. One finds
\begin{equation}
\label{eq21}
\begin{split}
G^{(s)}(\omega )&=\sum_{n=1}^7 \frac{\sigma ^{(s,n)}}{\omega
-E_n^{(s)} +i\delta },
\\
 C^{(s)}(\omega )&=\pi \sum_{n=1}^7 \sigma
^{(s,n)}T_n^{(s)} \delta (\omega -E_n^{(s)} ),
\end{split}
\end{equation}
where $T_n^{(s)} =1+\tanh ( {\beta E_n^{(s)} /2} )$ and the
spectral density matrices $\sigma ^{(s,n)}$ are computed by means
of the formula
\begin{equation}
\label{eq23} \sigma _{\mu \nu }^{(s,n)} =\Omega _{\mu n}^{(s)}
\sum_{\lambda =1}^7 [\Omega _{n\lambda }^{(s)} ]^{-1}I_{\lambda
\nu }^{(s)},
\end{equation}
where $\Omega ^{(s)}$ is the $7\times 7$ matrix whose columns are
the eigenvectors of the matrix $\varepsilon ^{(s)}$. The explicit
expressions of the spectral density matrices are given in the
appendix. $I^{(s)}$ is the normalization matrix defined as
\begin{equation}
\label{eq24} I^{(s)}=\langle \{\psi ^{(s)}(i),{\psi ^{(s)}}^\dag
(i)\}\rangle.
\end{equation}
By means of the recurrence relations (\ref{eq10}), all the matrix
elements of $I^{(s)}$ can be expressed in terms of only the
elements belonging to the first row. The calculations of the
latter gives, for a homogeneous state,
\begin{equation}
\label{eq25}
\begin{split}
I_{1,k}^{(\xi )} &=\kappa ^{(k-1)}-\lambda ^{(k-1)}\quad \quad
(k=1,...5),
 \\
I_{1,k}^{(\xi )} &=\delta ^{(k-5)}-\theta ^{(k-5)}\quad \quad
(k=6,7),
\\
I_{1,k}^{(\eta )} &=\lambda ^{(k-1)}\quad \quad (k=1,...5) , \\
I_{1,k}^{(\eta )} &=\theta ^{(k-5)}\quad \quad (k=6,7),
\end{split}
\end{equation}
where
\begin{equation}
\label{eq26}
\begin{split}
\kappa ^{(p)}&=\langle [n^\alpha (i)]^p\rangle, \\
\lambda ^{(p)}&=\frac{1}{2}\langle n(i)[n^\alpha (i)]^p\rangle,
\end{split}
\quad \quad
\begin{split}
\delta ^{(p)}&=\langle [D^\alpha (i)]^p\rangle, \\
\theta ^{(p)}&=\frac{1}{2}\langle n(i)[D^\alpha (i)]^p\rangle.
\end{split}
\end{equation}
In conclusion, in this Section we have shown that the 1D BEG model
is exactly solvable. Exact expressions for the GF and CF have been
obtained and are expressed in terms of a set of local correlation
functions (\ref{eq26}), which must be calculated in order to
obtain quantitative results. This problem will be considered in
the next Section, where a self-consistent scheme, capable to
compute the internal parameters, will be formulated.

\section{Self-consistent equations}
\label{sec_III}

On the basis of the computational framework provided in the
previous Section, it is evident that the GF and the CF depend on
the internal parameters: $\mu$, $\kappa ^{(p)}$ and $\lambda
^{(p)}$ ($p=0,\ldots ,4$), $\delta ^{(p)}$ and $\theta ^{(p)}$
($p=1,2$). For a homogeneous state (i.e., translationally
invariant: $\langle n^\alpha (i)\rangle=\langle n(i)\rangle$ and
$\langle D(i)\rangle=\langle D^\alpha (i) \rangle)$, there are
twelve parameters to be self-consistently computed in terms of the
external parameters $n$, $V$, $U$, $W$ and $T$. A first set of
self-consistent equations is given by the algebra constraints
\begin{equation}
\label{eq27}
\begin{split}
\xi _\uparrow (i)\xi _\uparrow ^\dag (i)&+\eta _\uparrow
(i)\eta_\uparrow
^\dag (i)=1-n_\uparrow (i), \\
\xi _\downarrow (i)\xi _\downarrow ^\dag (i)&+\eta _\downarrow
(i)\eta_\downarrow ^\dag (i)=1-n_\downarrow (i),
\\
\eta _\uparrow (i)\eta _\uparrow ^\dag (i)&=n_\downarrow (i)-D(i), \\
\eta _\downarrow (i)\eta _\downarrow ^\dag (i)&=n_\uparrow
(i)-D(i),
\end{split}
\end{equation}
from which one gets the following self-consistent equations
\begin{equation}
\label{eq28}
\begin{split}
 C_{1,1}^{(\eta )} &=\lambda ^{(0)}-\delta ^{(1)} ,\\
 C_{1,k}^{(\xi )} +C_{1,k}^{(\eta )} &=\kappa ^{(k-1)}-\lambda ^{(k-1)}\quad
(k=1,..5) ,\\
 C_{1,k}^{(\xi )} +C_{1,k}^{(\eta )} &=\delta ^{(k-5)}-\theta ^{(k-5)}\quad
(k=6,7),
 \end{split}
\end{equation}
where the CFs in the l.h.s. of Eq. (\ref{eq28}) can be computed by
means of the formula
\begin{equation}
\label{eq29} C^{(s)}=\langle \psi ^{(s)}(i){\psi ^{(s)}}^\dag
(i)\rangle=\frac{1}{2}{\kern 1pt}\sum_{n=1}^7 \sigma
^{(s,n)}T_n^{(s)}.
\end{equation}
Equations (\ref{eq28}) provide one with eight self-consistent
equations. To determine all the parameters one needs other four
equations. These can be derived by means of the algebra
constraints
\begin{equation}
\label{eq30} \xi ^\dag (i)n(i)=0 \quad \quad \xi ^\dag (i)D(i)=0.
\end{equation}
By exploiting these relations one can express the CFs
$C_{13}^{(\xi \xi )}$, $C_{14}^{(\xi \xi )}$, $C_{15}^{(\xi \xi
)}$, and $C_{17}^{(\xi \xi )}$ in terms of the CFs $C_{11}^{(\xi
\xi )}$, $C_{12}^{(\xi \xi )}$ and $C_{16}^{(\xi \xi )}$ as
\begin{equation}
\label{eq31}
\begin{split}
 C_{13}^{(\xi \xi )} &=C_{11}^{(\xi \xi )} \left(\frac{1}{2}X_1 +X_2
+\frac{1}{2}X_1 ^2 \right) ,\\
 C_{14}^{(\xi \xi )} &=C_{11}^{(\xi \xi )} \left(\frac{1}{4}X_1 +\frac{3}{2}X_2
+\frac{3}{2}X_1 X_2 +\frac{3}{4}X_1 ^2 \right), \\
 C_{15}^{(\xi \xi )} &=C_{11}^{(\xi \xi )} \left(\frac{1}{8}X_1 +\frac{7}{4}X_2
+\frac{9}{2}X_1 X_2 +\frac{7}{8}X_1 ^2+\frac{3}{2}X_2 ^2 \right), \\
 C_{17}^{(\xi \xi )} &=C_{11}^{(\xi \xi )} \left(\frac{1}{2}X_2 +\frac{1}{2}X_2^2
\right).
\end{split}
\end{equation}
The two parameters $X_1$ and $X_2$ are expressed in terms of the
CFs $C_{11}^{(\xi \xi )}$, $C_{12}^{(\xi \xi )}$ and $C_{16}^{(\xi
\xi )}$ as
\begin{equation}
\label{eq32} X_1 =\frac{C_{12}^{(\xi \xi )} }{C_{11}^{(\xi \xi )}
}, \quad \quad  X_2 =\frac{C_{16}^{(\xi \xi )} }{C_{11}^{(\xi \xi
)} }.
\end{equation}
Equations (\ref{eq28}), (\ref{eq31}) and (\ref{eq32}) provide
twelve self-consistent equations which will determine all the
unknown internal parameters and therefore the various properties
of the model. Once the parameters of the fermionic model are
computed, by use of the mapping transformations \eqref{eq4} and
\eqref{eq5}, it is straightforward to study the behavior of
relevant properties of the BEG model. Details of the computations
leading to Eq. \eqref{eq31} will be given elsewhere
\cite{mancini2}.

\section{Magnetic and Thermal Responses}
\label{sec_IV}

As an application of the general formulation provided in the
previous sections, here we shall study the magnetic and thermal
properties of the model, by restricting the analysis to the case
$K=0$ and $J<0$.
\begin{figure}[t]
\centerline{\includegraphics[scale=0.8]{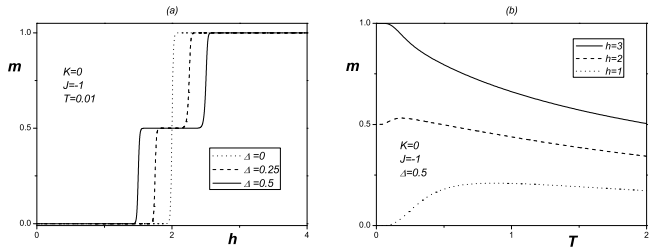}}
 \caption{\label{fig1} (a) The magnetization $m$ as a function of the
 external field  $h$ for $K=0$, $J=-1$, $T=0.01$ and for  $\Delta=0, 0.5$ and
 1. (b) The magnetization as a function of the temperature for $K=0$, $J=-1$, $\Delta=0.5$ and $h=1$, 2 and 3.}
 \end{figure}
In the following we set $J=-1$ and we consider only positive
values of $h$, owing to the symmetry property of the model under
the transformation $h \to -h$. When $\Delta=0$ the ground state is
either antiferromagnetic or ferromagnetic, depending on the value
of the external field. As a consequence, the magnetization,
defined as
\begin{equation}
m=\langle S(i)\rangle=\langle n(i)\rangle-1=1-2[C_{11}^{(\xi )}
+C_{11}^{(\eta )} ],
\end{equation}
presents at $T=0$ two plateaus as a function of the external field
$h$, with $h_s=2$ being the value of the saturated field. Turning
on a positive single-ion anisotropy, three plateaus appear at
$m=0$, $m=1/2$ and $m=1$. This is in agreement with the criterion
derived in Ref. \cite{oshikawa} for the appearance of plateaus in
spin chains in a uniform magnetic field.

In Fig. \ref{fig1}a we plot the magnetization as a function of the
magnetic field at $T=0.01$ for different values of $\Delta$. Upon
increasing the field, a nonzero magnetization begins at the
critical value of the field $h_c$: this critical value decreases
by increasing $\Delta$. For $h>h_c$ a $m=1/2$ plateau is observed
until $h$ reaches the saturated value $h_s$, at which the third
magnetization plateau at $m=1$ is observed. The width of the
$m=1/2$ plateau augments by increasing $\Delta$ in the range
$0<\Delta<1$ and becomes independent of $\Delta$ when $\Delta>1$.
The critical field $h_c$ and the saturated field $h_s$ satisfy, in
the range $0<\Delta<1$, the laws:
\begin{equation*}
h_c = 2 -\Delta, \qquad \quad h_s=2+\Delta .
\end{equation*}
Our findings are in good agreement with Monte Carlo \cite{chen}
and transfer matrix \cite{aydiner} results. In Fig. \ref{fig1}b we
plot the magnetization as a function of the temperature for values
of the magnetic field belonging to the three different plateaus.
For $h=1$ ($h<h_c$) the magnetization is zero at $T=0$; all spins
are aligned (upward and downward) with the magnetic field,
resulting in a pure AF state. When the temperature increases, the
thermal fluctuations allow some of the downward spins to rotate
and the magnetization increases up to $T \approx 0.9$  where it
exhibits a maximum. Further increasing $T$, the thermal
fluctuations enter in competition with the magnetic field and the
magnetization decreases. For $h=2$ (intermediate phase
$h_c<h<h_s$) the magnetization is equal to 1/2 at $T=0$: half of
the spins are parallel to $h$ and half lie in the transverse
plane. When the temperature increases, there is a slight increase
of $m$ (up to $T \approx 0.2$), but soon the disorder induced by
thermal fluctuation prevails and $m$ decreases. For $h=3$
($h>h_s$) at $T=0$ one finds $m=1$: all spins are parallel to the
magnetic field and the system is in a pure ferromagnetic state.
When the temperature increases, the long-range order is destroyed
and the magnetization decreases.

\begin{figure}[t]
\centerline{\includegraphics[scale=0.8]{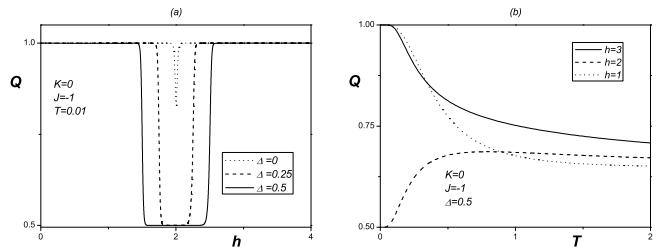}}
 \caption{\label{fig2} (a) The quadrupolar moment $Q$ as a function of the
 external magnetic field $h$ for $K=0$, $J=-1$, $T=0.01$ and for  $\Delta=0, 0.5$ and
 1. (b) The quadrupolar moment as a function of the temperature for $K=0$, $J=-1$, $\Delta=0.5$ and $h=1$, 2 and 3.}
 \end{figure}

To further analyze the magnetic behavior, we have studied the
quadrupolar moment $Q$, defined as
\begin{equation}
Q=\langle S^2(i)\rangle=1-2C_{11}^{(\eta )}.
\end{equation}
At zero temperature also this quantity shows plateaus for $\Delta
\ge 0$. In Fig. \ref{fig2}a the quadrupolar moment $Q$ is plotted
as a function of the external magnetic field, for $K=0$, $J=-1$,
$T=0.01$ and for various values of $\Delta$. $Q$ takes the value
$1/2$ in the range $h_c < h < h_s$, whereas is equal to 1 for all
other values of $h$. The behavior of $Q$ as a function of the
temperature is shown in Fig. \ref{fig2}b. For $h<h_c$ and $h>h_s$,
the quadrupolar moment $Q$ is maximum ($Q=1$) at $T=0$ and
decreases by increasing $T$. For $h_c<h<h_s$, $Q$ vanishes at zero
temperature and increases augmenting $T$.

\begin{figure}[t]
\centerline{\includegraphics[scale=0.8]{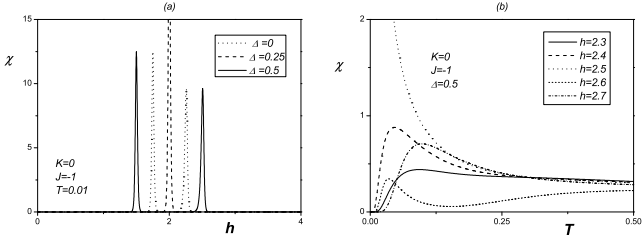}}
 \caption{\label{fig3} (a) The susceptibility $\chi$ as a function of the
external field for $K=0$, $J=-1$, $T=0.01$ and for $\Delta=0$, 0.5
and 1. (b) The susceptibility as a function of the temperature for
$K=0$, $J=-1$, $\Delta=0.5$ in the neighborhood of $h_s$.}
 \end{figure}

The existence of the magnetic plateaus is endorsed by the peaks
found in the magnetic susceptibility $\chi= dm/dh$. As evidenced
in Fig. \ref{fig3}a, for $\Delta =0$ one finds only one peak,
whereas for $\Delta>0$ there are two peaks appearing at $h_c$ and
$h_s$, signalling a step-like behavior of the magnetization. When
plotted as a function of the temperature, $\chi$ shows a peak at
low temperatures and then vanishes for $T \to 0$ for all values of
the magnetic field but at $h_{c,s}$, where, of course, it diverges
due to the step encountered by the magnetization. In Fig.
\ref{fig3}b, we plot, as an example, the susceptibility in the
neighborhood of $h_s$.

The specific heat is given by $C=dE/ dT$ where the internal energy
$E$ can be computed as the thermal average of the Hamiltonian
\eqref{eq1} for $K=0$
\begin{equation}
E=-J \langle S(i) S(i)^{\alpha} \rangle + \Delta \, Q-h \, m .
\end{equation}

\begin{figure}[t]
\centerline{\includegraphics[scale=0.8]{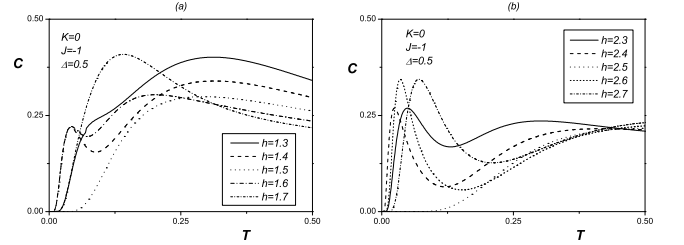}}
 \caption{\label{fig4} (a) The specific heat  $C$ as a function of the temperature
 for $K=0$, $J=-1$,  $\Delta=0.5$ and
 in the neighborhood of $h_c$. (b)
 The specific heat as a function of the temperature for $K=0$, $J=-1$,  $\Delta=0.5$ and
 in the neighborhood of $h_s$.}
 \end{figure}

The specific heat exhibits a rich structure in correspondence of
the critical values of the magnetic field. The behavior of the
specific heat as a function of the temperature is shown in Figs.
\ref{fig4}a-b in the neighborhood of $h_{c,s}$. The possible
excitations of the ground state are flipping of the spins parallel
to $h$ and/or of the ones perpendicular to it. Far from the
critical values, the specific heat presents only one peak at low
temperatures. When $h<h_c$, the possible excitations are due only
to the flipping of the longitudinal spins. In the neighborhood of
$h_c$, a second peak appears since thermal fluctuations tends also
to flip the spins perpendicular to the external field. Further
increasing $h$, the specific heat presents only one peak until $h
\approx h_s$, where again two peaks are present in a broader
region with respect to $h_c$. Away from $h_s$, the specific heat
shows only one peak.

\section{Concluding Remarks}
\label{sec_V}

We have evidenced how the use of the Green's function and
equations of motion formalism leads to the exact solution of the
one-dimensional BEG model. Our analysis allows for a comprehensive
study of the model in the whole space of parameters $K$, $J$,
$\Delta$, $h$ and $T$. Here, we have focused on the
antiferromagnetic properties exhibited by the model and we have
shown that, at zero temperature, the model exhibits three magnetic
plateaus when $\Delta >0$. Furthermore, the specific heat shows a
double peak structure in the neighborhood of the endpoints of the
intermediate plateau.

\section*{Acknowledgments}

This paper is dedicated to Professor Ihor Stasyuk on the occasion
of his 70th birthday, wishing him many more years of successful
and fruitful work.

\appendix

\section{Some analytical expressions}
The coefficients $A_m^{(p)} $ and $B_m^{(p)}$ in Eq. \eqref{eq10}
are given by:
\begin{equation}
\label{eq12}
\begin{split}
 A_1^{(p)} &=-6+2^{3-p}-2^{p-1}+2^{3-p}\cdot 3^{p-1} \\
 A_2^{(p)} &=\frac{-104+57\cdot 2^{p+1}-56\cdot
3^p+11\cdot 4^p}{3\cdot 2^{p+1}} \\
 A_3^{(p)} &=\frac{18-3\cdot 2^{p+3}+14\cdot 3^p-3\cdot
4^p}{3\cdot 2^{p-1}} \\
 A_4^{(p)} &=\frac{-4+3\cdot 2^{p+1}-4\cdot 3^p+4^p}{3\cdot 2^{p-1}}
 \end{split}
\end{equation}
and
\begin{equation}
\label{eq13}
\begin{split}
 B_1^{(p)} &=2^{2-p}-1 \\
 B_2^{(p)} &=2-2^{2-p} .
 \end{split}
\end{equation}
$\sigma ^{(s,m)}=\Sigma_m^{(s)} \Gamma ^{(m)}$ are the spectral
density matrices defined in Eq. \eqref{eq23}; $\Gamma^{(m)}$ are
matrices of rank $7 \times 7$:
\begin{equation}
\begin{split}
\Gamma_{1,k}^{(1)} &= \scriptstyle \left(1 \; \, 0 \; \, 0 \; \,
0\; \, 0 \; \, 0 \; \, 0 \right)
 \\
\Gamma_{l,k}^{(2)} &= \scriptstyle \left( 1 \; \, 1/2 \; \, 1/4 \;
\,  1/8\; \, 1/16 \; \, 0 \; \, 0
\right) \\
\Gamma_{l,k}^{(3)} &=  \scriptstyle\left( 1 \; \, 1 \; \, 1 \; \,
1 \; \, 1 \; \, 0 \; \, 0 \right)
\\
\Gamma_{l,k}^{(4)} &=  \scriptstyle\left( 1 \; \, 1 \; \, 1 \; \,
1 \; \, 1 \; \, 1 \; \, 1 \right)
\\
\Gamma_{l,k}^{(5)} &= \scriptstyle \left( 1 \; \, 1 \; \, 1 \; \,
1 \; \, 1 \; \, 1/2 \; \, 1/4 \right)
\\
\Gamma_{l,k}^{(6)} &=  \scriptstyle\left( 1 \; \, 2 \; \, 4 \; \,
8\; \, 16 \; \, 1 \; \, 1
\right) \\
\Gamma_{l,k}^{(7)} &=  \scriptstyle \left( 1 \; \, 3/2 \; \,8/4 \;
\, 27/8\;\, 81/16 \;\, 1/2 \;\, 1/4 \right).
\end{split}
\end{equation}
and the $\Sigma _m^{(s)}$ are given by:
\begin{equation*}
\begin{split}
\Sigma_1^{(s)} &=  \scriptstyle \frac{1}{6}\left(6I_{1,1}^{(s)}
-25I_{1,2}^{(s)} +35 I_{1,3}^{(s)} -20I_{1,4}^{(s)}
+4I_{1,5}^{(s)} \right)
\\
\Sigma_2^{(s)} &=  \scriptstyle \frac{4}{3}\left(6I_{1,2}^{(s)}
-13I_{1,3}^{(s)} +9I_{1,4}^{(s)} -2I_{1,5}^{(s)} \right)
 \\
 \Sigma_3^{(s)}
 &=  \scriptstyle -\frac{23}{6}I_{1,2}^{(s)}+\frac{23}{2}I_{1,3}^{(s)}-\frac{26}{3}
 I_{1,4}^{(s)} + 2I_{1,5}^{(s)}-3I_{1,6}^{(s)}+ 2I_{1,7}^{(s)}
 \\
\Sigma_4^{(s)} &=  \scriptstyle   \frac{1}{6}\left(3I_{1,2}^{(s)}
-11I_{1,3}^{(s)}
+12I_{1,4}^{(s)}-4I_{1,5}^{(s)}-6I_{1,6}^{(s)}+12I_{1,7}^{(s)}
\right)
\end{split}
\label{A_13}
\end{equation*}
\begin{equation*}
\begin{split}
 \Sigma_5^{(s)} &=  \scriptstyle\frac{4}{3}\left(-2I_{1,2}^{(s)} +7I_{1,3}^{(s)}
-7I_{1,4}^{(s)} +2I_{1,5}^{(s)} +3I_{1,6}^{(s)}-3I_{1,7}^{(s)}
\right)
 \\
 \Sigma_6^{(s)} &= \scriptstyle \frac{1}{6}\left(-3I_{1,2}^{(s)} +11I_{1,3}^{(s)}
-12I_{1,4}^{(s)} +4I_{1,5}^{(s)} \right)
\\
 \Sigma_7^{(s)} &= \scriptstyle \frac{4}{3}\left(2I_{1,2}^{(s)}-7I_{1,3}^{(s)}
+7I_{1,4}^{(s)} -2I_{1,5}^{(s)} \right).
\end{split}
\label{A_14}
\end{equation*}
Here we have reported only the first row of the spectral density
matrices. All the other matrix elements can be expressed in terms
of the first row by means of the recursion relation \eqref{eq10}.

\end{document}